# Time-modulated Hamiltonian for interpreting Mach-Zehnder interferometer delayed-choice experiments

Zhi-Yuan Li

Laboratory of Optical Physics, Institute of Physics, Chinese Academy of Sciences, Beijing 100190, China and College of Physics and Optoelectronics, South China University of Technology, Guangzhou 510641, China

Email address: lizy@aphy.iphy.ac.cn

**Abstract**

Many delayed-choice experiments based on Mach-Zehnder interferometers (MZI) have been thought and made to address the fundamental problem of wave-particle duality. Conventional wisdoms long hold that by inserting or removing the second beam splitter (BS2) in a controllable way, microscopic particles (photons, electrons, etc.) transporting within the MZI can lie in the quantum superposition of the wave and particle state as $\psi = a_w \psi_{wave} + a_p \psi_{particle}$. Here we present an alternative interpretation to these delayed-choice experiments. We notice that as all composite devices of MZI including BS2 are purely classical, the inserting and removing operation of BS2 imposes a time-modulated Hamiltonian $H_{mod}(t) = a(t) H_{in} + b(t) H_{out}$, instead of a quantum superposition of $H_{in}$ and $H_{out}$ as $H = a_w H_{in} + a_p H_{out}$, to act upon the incident wave function. Solution of this quantum scattering problem, rather than the long held quantum eigen-problem yields a synchronically time-modulated output wave function as $\psi_{mod}(t) = a(t) \psi_{wave} + b(t) \psi_{particle}$. As a result, the probability of particle output from the MZI behaves as if they are in the superposition of the wave and particle state when many events over time accumulation are counted and averaged. We expect these elementary but insightful analyses will shed a new light on exploring basic physics beyond the long-held wisdom of wave-particle duality and principle of complementarity.

# I. Introduction

The interpretation of quantum theory, a conceptual foundation problem in physics, has been an issue ever since its founding nearly 100 years ago [1-11], although the operation power of quantum mechanics to solve practical problems in microscopic world is of no dispute. This conceptual foundation has raised extensive hot controversies in history, e.g., between Einstein and Bohr [1-3,5], and is still attracting much attention in current days and raising intensive disputes and discussions. The wave-particle duality of quantum objects, and more generally, the principle of complementarity, stands on the central conceptual core of quantum theory. As Feynman famously stated, the wave-particle duality as illustrated in double-slit experiment has in it the heart of quantum mechanics; in reality it contains the only mystery of the theory [8].

According to the orthodox Copenhagen interpretation, all quantum objects including massless and massive particles exhibit mutually exclusive behaviors of two intrinsic attributes of the wave nature and particle nature, namely, they behave either as wave or as particles, depending on how they are observed and measured, but never both. In history, numerous studies, either theoretical or experimental, have been made to test the wave-particle duality aiming to gain deeper understanding on the conceptual foundation of quantum mechanics [12]. Most experiments favored the orthodox Copenhagen interpretation that one can never observe simultaneously the wave and particle behavior of quantum objects using a single set of experimental apparatus, even in theory and in principle. In this paper we revisit this historical problem by analyzing quantitatively delayed-choice experiments in the framework of orthodox quantum mechanics operation formulation and show that the result will shed a new light on understanding of this basic problem.

# II. Principle of Mach-Zehnder Interferometer

A popular instrument to demonstrate and analyze the puzzling feature of wave-particle duality is the Mach-Zehnder interferometer (MZI). A standard form of MZI is schematically illustrated in Fig. 1, which involves several key components.

The first beam splitter (BS1) is used to split the incident coherent particle beam (either massless particles such as photons or massive particles such as electrons and neutrons) into two equal-weight particle beam, assuming a 50:50 performance. Two mirrors (M) are used to displace and adjust the direction of transport path of the particle beam. A phase shift plate is used to introduce adjustable path and phase difference $\varphi$ between the two arms (path *X* and path *Y*) of MZI. The second beam splitter (BS2), again assuming a 50:50 performance, is used to combine particle beams from the two arms into the individual output port connected with detector *x* and detector *y*, respectively.

It is well known that at the presence of both BS1 and BS2, this MZI can perfectly demonstrate the wave nature of quantum particles. When a high-coherence particle beam is sent into this MZI, the signal intensity measured by the detector *x* and *y* in this MZI exhibits a perfect cosine function of the phase difference $\varphi$ with a well-defined, in principle infinitely fine signal peak-to-valley contrast. In mathematics, the signal is given by $P_{y(x)}(\varphi) = |\psi_0|^2 [1 \pm \cos\varphi]/2$. The presence of perfect interference fringe curve is an ideal representation of the wave nature of the quantum particle. In the orthodox conceptual formulation of quantum theory, this result originates from the fact that it is now completely (100%) impossible to identify the transport path (either path *X* or path *Y*) of each particle detected by the two detectors *x* and *y*, or in other words, the path information of each particle is completely lost.

On the other hand, if one wants to identify unanimously which path the particle transports, then a simple modification can be made to the old MZI, namely, removing BS2 from the setup. The path of the particle can be unambiguously determined by looking at the signal recorded by detector *x* and *y* in this new MZI, as schematically illustrated in Fig. 2. Yet, the price is that now one completely loses the capability to observe the wave nature of the particle because the interference fringe with respect to the phase difference $\varphi$ is 100% smeared out and leaves a constant line in both detector *x* and *y*. Mathematically the signal is $P_{y(x)}(\varphi) = |\psi_0|^2/2$. Obviously the old and new version of the MZI each cannot simultaneously tell the wave and particle nature of the quantum object, instead they only tell either one or the other nature (i.e., wave or particle).

This is the classical story of wave-particle duality test by using MZI, and a perfect example to illustrate the principle of complementarity. Yet, the desire of physicists to fully uncover the secret behind the motion of quantum particle ignites flashes of wisdom from time to time. The delayed-choice thought experiments as first proposed by Wheeler in late 1970s are among such kinds of prominent examples [13-18]. In these schemes, people try to use some very smart conceptual arguments or state-of-the-art experimental technologies to resolve the puzzles of wave-particle duality for each single quantum particle by observing their motion in various MZIs. Such studies would allow people to have deeper insights into the delicate behavior of motion of quantum particles and hopefully might discover the laws under these motions that go beyond the orthodox quantum theory as systematically presented in the framework of the Schrödinger equation and wave function, together with its conceptual orthodox Copenhagen interpretation.

A classic example of delayed-choice experiment is illustrated in Fig. 3 based on the MZI. The BS2 is controlled by external operation to either go into the interferometer, called as the "in" state, or come out of the interferometer, called as the "out" state. However, the classical wave-particle duality firmly stands in these smart thought or practical experiments. The wave and particle nature of quantum particles are still exclusively repulsive to each other and nobody is able to observe simultaneously the wave and particle nature of quantum particles. More recently, several works have moved one further step and gone into the regime of so-called quantum delayed-choice experiments by replacing the classical BS2 with a quantum BS2 device that can lie in the quantum superposition state of the "in" and "out" state [19-25]. Theoretical and experimental results show that the output of the MZI exhibits the coexistence (or mixture) of partial wave nature and partial particle nature of quantum particles. Nonetheless, these quantum delayed-choice experiments are still not able to observe the full wave nature and particle nature simultaneously and thus the results do not violate the principle of complementarity.

**III. Quantum Scattering Problem Analysis**

It is worth noting that although the problem in Figs. 1-3 is far from complicated, a thorough and systematic quantum-mechanical analysis over them has been very rare in history, perhaps because they seem to be too simple and thus a simple conceptual argument is sufficient to find the correct answer. In this section we follow an alternative path to make an elementary while insightful analysis of these MZI experimental schemes used to test the wave-particle duality, hoping to shed a new light on this old problem. Our analysis is strictly based on the formalism of classical quantum mechanics. Although the conceptual basis of quantum theory is in dispute for a long time, the operation formalism of quantum theory and its success and power has never been a controversial issue. Our analyses turn out to show that quantum mechanics of the current MZI problem involves several key ingredients, however, the most important thing of all is that it is a quantum scattering problem, rather than the usual quantum mechanical eigen-problem that most literatures have exclusively adopted until now. This means that the output physical quantity measured and analyzed by detectors is the quantum scattering amplitude or intensity of a particle beam passing through the MZI, a quantum device. This output should closely rely on the quantum state of the input beam, and can be calculated precisely if strict quantum mechanical analysis and solution are made.

The quantum mechanics for the MZI satisfies the following Schrödinger equation

$$i\hbar \frac{\partial}{\partial t}\Psi(R,t) = H(R,t)\Psi(R,t). \qquad (1)$$

Here $H(R,t)$ is the Hamiltonian of the MZI, which should rely on the specific geometric and physical configuration (denoted as $R$) of all the composite devices of the MZI as enclosed within the dashed rectangular box in Figs. 1-3, including the two mirrors, two beam splitters, and the phase plate. In most MZI experiments one can straightforwardly write down the explicit form of this Hamiltonian, and usually it is an elastic scattering Hamiltonian where the energy $E$ of the quantum particle is conserved. $\Psi(R,t)$ is the wave function of the quantum particle that transports within the MZI. Since this is a quantum scattering problem, the determination of $\Psi(R,t)$ should closely rely on the boundary condition (at $R_\infty$) as well as the initial condition (at $t_0$), which is written as the input wave function $\Psi_{input}$ at the entrance port of the MZI. At the exit port of the MZI, the wave function is $\Psi_{output}$, which is the quantity that the detector $x$ and $y$ probe and analyze for testing the wave-particle duality.

Given the explicit form of both $H(R,t)$ and $\Psi_{input}$, the unknown wave function $\Psi_{output}$ can be readily calculated. For photons, the analysis is about electromagnetic wave (representing photons). The methodology of analysis has long been well established and can be found in classical optics textbooks. It eventually traces back to the solution of Maxwell's equations for photon transporting through all the optical devices of the MZI, which are purely classical in nature. For massive particles such as electrons one needs to solve the Schrödinger equation for these particles transporting through all the composite devices of MZI. In essence, one can readily follow the path of the quantum particle and deal with the transmission and reflection of wave function at each device one by one.

Now that the general framework of analyzing the quantum mechanical problem of MZI has been clarified, we proceed to see what happens in Figs. 1-3. The MZI setup used to illustrate the wave nature of a quantum particle, with the BS2 present, has already been depicted in Fig. 1, where the Hamiltonian of the setup is denoted as $H_{in}(R)$. Here the subscript "in" means the "in" state of BS2. Note that because this Hamiltonian is time independent (or stationary), the scattering problem has a stationary solution, $\Psi_{in}(R,t) = \Psi_{in}(R)e^{-iEt/\hbar}$, where $E$ is the energy (more precisely the kinetic energy) of the particle. The signal intensity is $P_{in} = |\Psi_{in}(R,t)|^2 = |\Psi_{in}(R)|^2$ and also time independent.

Assume that the incident particle beam is described by a wave function $\Psi_{input} = \psi(\mathbf{r}, x) = \psi_0(\mathbf{r})e^{ikx}$, where $\mathbf{r}$ and $x$ denote the transverse and longitudinal space coordinate of the particle beam. Thus $\psi_0(\mathbf{r})$ and $e^{ikx}$ represent the transverse and longitudinal freedom of the particle beam wave function. In many experiments, $\psi_0(\mathbf{r}) = \psi_0$ is a constant, thus $\Psi_{input}$ is a plane wave transporting along the $x$-axis, i.e., the horizontal direction. A simple classical-optics like calculation shows that the solution of this MZI quantum mechanical scattering problem at the $x$ output port is

$$\Psi_{output,x} = \Psi_{in,x} = \psi_0(1 - e^{i\varphi})e^{ikx} / 2. \tag{2}$$

The corresponding signal intensity is

$$P_{in} = |\Psi_{in,x}|^2 = |\psi_0|^2 [1 - \cos\varphi] / 2 = P_0[1 - \cos\varphi]. \tag{3}$$

Here $P_0 = |\psi_0|^2 / 2$ is the signal intensity at each path. As illustrated in Fig. 1, the signal shows periodic oscillation with respect to the change of phase shift $\varphi$, which is a perfect interference pattern with the fringe visibility $V^2 = 1$ [26]. One can judge from this feature that the quantum particle behaves completely as a wave, and thus the corresponding quantum state can be designated as $\Psi_{in} = \psi_{wave}$. Yet, the path information of the particle is completely lost now with the path distinguishability being $D^2 = 0$ [26].

Now let us look at Fig. 2, where BS2 is removed from the MZI. The corresponding time-independent Hamiltonian and wave function are denoted as $H_{out}(R)$ and $\Psi_{out}(R,t) = \Psi_{out}(R)e^{-iEt/\hbar}$, respectively. Here the subscript "out" means the "out" state of BS2. The solution of the output wave function under a given input wave function $\Psi_{input} = \psi(\mathbf{r}, x) = \psi_0(\mathbf{r})e^{ikx} = \psi_0 e^{ikx}$ can be calculated, which is

$$\Psi_{output,x} = \Psi_{out,x} = \psi_0 e^{i\varphi} e^{ikx} / \sqrt{2}. \tag{4}$$

The signal intensity is

$$P_{out} = |\Psi_{out,x}|^2 = |\psi_0|^2 / 2 = P_0. \tag{5}$$

As illustrated in Fig. 2, the signal is just a constant irrespective of the phase shift $\varphi$, and with a fringe visibility $V^2 = 0$, it does not show any wave nature. Yet, the path information of the quantum particle can be unanimously identified now, and the corresponding path distinguishability is $D^2 = 1$. The corresponding quantum state can be designated as $\Psi_{out} = \psi_{particle}$. The above two simple quantum mechanical calculations confirm the well-known results in history about the MZI used to test the wave-particle duality.

We proceed to look at the more complicated situation as shown in Fig. 3. The status of BS2 is externally controlled to move into or out of the MZI, which can be random or other forms. The Hamiltonian of this new system is thus time dependent, denoted as $H_{mod}(R,t)$. The corresponding wave function should be much more complicated than the stationary solution $\Psi(R,t) = \Psi(R)e^{-iEt/\hbar}$. Yet, in most practical situations, some approximations can be adopted, rendering easy solution of the quantum mechanical problem.

One prominent point in Fig. 3 is that two sets of instrumental setup, which are still all purely classical devices, are used, namely, the setup with BS2 inserted or removed, respectively. These two setups cannot be in place simultaneously in time, because they obviously repel each other. In the methodology of quantum theory, these two sets of instrument arrangement correspond to two mutually repulsive quantum mechanical problems corresponding to the two different setups of MZI as illustrated in Figs. 1 and 2, respectively. Their quantum mechanical solution, as has been discussed in Eqs. (2)-(5), only allows for either wave or particle behavior observation of quantum particles. The current scheme of Fig. 3 uses a single set of measurement setup involving two instruments placed in different space-time domains, in the form of some deliberate temporal operation and modulation to BS2. The question is, can this clever delayed-choice experimental setup allow for uncovering new physics, say, simultaneous observation of the wave and particle nature of quantum objects, without offending any standard methodology of quantum theory?

Keeping in mind the above basic physical picture, the Hamiltonian in Eq. (1) is written into a time-modulated form as

$$H_{mod}(R,t) = a(t)H_{in}(R) + b(t)H_{out}(R), \tag{6}$$

where $a(t)$ and $b(t)$ are temporal function used to collectively describing how BS2 goes into (operation "in") and moves out of (operation "out") the MZI. Usually they are step function given by $a(t)=1, b(t)=0, t \subset [t_1,t_2]$ or $a(t)=0, b(t)=1, t \subset [t_1,t_2]$. This simple expression means that the two operations cannot happen simultaneously at any time interval $[t_1,t_2]$. Besides, the time series $[t_1,t_2]$ and duration $\Delta t = t_2 - t_1$ can be of arbitrary form, depending on how the state of BS2 is controlled and modulated.

Suppose that in practical experiments the modulation time $\Delta t$, i.e., transition from the operation "in" to the operation "out', is far longer than the transport time of particle through BS2, then the transient behavior of the wave function at this transition window can be neglected when the overall evolution dynamics of the particle is concerned. Keeping this in mind, we find the output wave function from the MZI as detected and analyzed by the detectors $x$ and $y$ can be expressed as

$$\Psi_{output}(R,t) = \Psi_{mod}(R,t) = a(t)\Psi_{in}(R,t) + b(t)\Psi_{out}(R,t). \tag{7}$$

Thus the output wave function is simply a superposition of the wave behavior state and particle behavior state in temporal domain, call temporal superposition, in contrast to the usual superposition of stationary states in quantum mechanics. When one follows the regular practice in carrying out this kind of quantum mechanical experiment, and accumulate in time and average over the signal recorded and analyzed by the detectors, the expected outcome of signal intensity is given by

$$P_{output} = P_{mod} = A|\Psi_{in}|^2 + B|\Psi_{out}|^2 = AP_{in} + BP_{out}. \tag{8}$$

The coefficient $A$ and $B$ are given by

$$A = \frac{1}{T}\int_0^T |a(t)|^2 \, dt, \quad B = \frac{1}{T}\int_0^T |b(t)|^2 \, dt. \tag{9}$$

Here $T$ is the total accumulation time of the signal. Obviously $A$ and $B$ are just the ratio of the total time interval of "in" and "out" operation state of BS2, respectively over the total accumulation time *T*. In all cases of temporal modulation, $A + B = 1$. Without losing any generality, we can define $A = \sin^2\theta$, $B = \cos^2\theta$, then the signal in the MZI scheme of Fig. 3 is calculated from Eq. (8) as

$$P_{output} = P_{mod} = P_0[1 - \sin^2\theta \cos\varphi]. \tag{10}$$

This equation can be written into another equivalent form as

$$P_{output} = P_{mod} = 2P_0[\cos^2\frac{\varphi}{2}\sin^2\theta + \frac{1}{2}\cos^2\theta]. \tag{11}$$

It is interesting to note that Eq. (11) is exactly the same as Eq. (6) in Ref. [21]. As illustrated in Fig. 3, the detected signal is just the superposition (or more precisely, the mixture) of the particle and wave state of quantum particle, namely, a perfect cosine function superimposed with a non-zero offset signal basis. It involves partial wave feature exhibiting an interference fringe with visibility of $V^2 = \sin^2\theta$ and partial particle feature exhibiting the which-path information with path distinguishability $D^2 = \cos^2\theta$. Obviously the outcome of signal can continuously transform from the purely wave state ($\theta = 0$) to the purely particle state ($\theta = \pi/2$), and locate in any intermediate superposition (mixture) state between them. In any case, we find $V^2 + D^2 = 1$, thus this delayed-choice experimental scheme is impossible for simultaneous observation of perfect wave and particle nature of quantum particle, which requests $V^2 = D^2 = 1$ and $V^2 + D^2 = 2 \gg 1$ in ideal conditions [19-27].

## IV. Comparison with Previous Analyses

Our analysis made in the above sections involves several key points. First, the analysis completely and strictly follows the operation formulation of standard quantum mechanics (the Schrödinger equation and wave function, together with boundary and initial conditions), which are well-established, have good consensus, and are free from controversies. Second, the analysis uses the probability interpretation of wave function against practical experimental observations, namely, any expected result of measurement over a physical quantity against a quantum state is an average over the outcome of many identical measurement operations against that specific quantum state, and this is again free from controversy. In the current situation of MZI, the outcome of wave nature (the interference pattern) and particle nature (which-way information) measurement is determined via accumulation and average in time over the detector outcome when the MZI experiment is performed over single particle one by one in time. Thus we justify that our analysis is made on a solid basis of quantum mechanics rather than pure conceptual arguments.

On the other hand, our analysis also yields useful conceptual hints to better understand the physics of MZI experiments made either in history or in modern time. First, all devices in the MZI are classical and their masses are many orders of magnitude larger than the quantum particles under study, thus, their response under external action and operation, e.g., the "in" and "out" operation of BS2, is purely classical. This is still the situation in recent several quantum delayed-choice experiments [19-25]. Second, the interaction of these devices with the particles and the consequent modulation to their wave function must be quantum mechanically described and solved. Third, the counteraction of the particle to these classical devices can be completely omitted. As a result, the quantum mechanical problem of particle transport and wave function evolution through the MZI becomes a simple and straightforward single-particle quantum elastic scattering problem. Finally, the detectors in the MZI are designed and managed only to count the single-particle events and thus monitor the single-particle quantum state, either wave state or particle state. They have nothing to do with the two-particle or even many-particle events and quantum states (usually recorded and evaluated via coincidence counting technique). Also notice that although in principle one can do so, as has been done in many recent experiments [19-25], but essentially there is no need to do so, because what concerns is the wave-particle duality of a quantum particle, which is intrinsically a single-particle feature.

As the only way the traditional delayed-choice experiment designers can do is to change the status of BS2 in order to test the wave-particle duality via the MZI in

Fig. 3, there is no way that this MZI instrument can tell simultaneously the wave and particle nature of quantum particles according to our above analysis. The measurement outcome of the detectors has three options. It either tells unambiguously the wave nature of the quantum particle by permanently inserting BS2 in the MZI, or tells unambiguously the wave nature of the quantum particle by permanently removing BS2 from the MZI, or yields a result that seemingly matches with an intermediate superposition state between the wave and particle as intensively discussed in recent literatures of quantum delayed-choice experiments [19-25]. Surely there are new physics, but the fundamental cause is far from being completed.

At the first glance, our observation based on quantum mechanical analysis over a quantum scattering problem under a time-modulated Hamiltonian seems to be exactly the same with the prediction made by previous delayed-choice experiment analyses [19-25]. In particular, let's remind again the exact identification of Eq. (11) derived from our time-modulated Hamiltonian analysis to Eq. (6) in Ref. [21] derived by the classical quantum-delayed choice experiment analysis. Nonetheless, the physical picture is very much different between our analysis and previous analyses. Whereas the outcome of measurement over the which-path information and interference fringe visibility in this MZI as illustrated in Fig. 3 is the same in both analyses, the detailed physical process is very different. In our analysis, the wave-particle duality test is a quantum mechanical scattering problem. The motion of BS2 in the MZI induces a time-modulated Hamiltonian acting over the incident wave function of the quantum particle, leading to the output of a synchronously time-modulated quantum state and the seemingly coexistence and mixture of partial wave and partial particle behavior. In previous analyses, the wave-particle duality test is a quantum mechanical eigen-problem. The motion of BS2 under external modulation action induces a new time-independent stationary quantum state, namely, the superposition state of wave state and particle state, $\psi = a_w \psi_{wave} + a_p \psi_{particle}$, where $a_w$ and $a_p$ are coefficients. Obviously this new quantum state carries both the information of wave and particle. The detectors directly interact and deal with this quantum state, and record the same outcome as our current analysis, namely, Eq. (10) [or Eq. (11)]. Frankly speaking, apart from conceptual differences, when only the technical issue is concerned, our analysis is much simpler and more straightforward than previous delayed-choice experiment analysis.

A simple argument shows that the new quantum state should be the solution of a stationary quantum superposition Hamiltonian as $H = a_w H_{in} + a_p H_{out}$. It is not evident from previous analyses how the classical device BS2 can stay in such a superposition state and how its Hamiltonian can be in such a superposition (or merger) of two completely exclusive Hamiltonian as $H_{in}$ and $H_{out}$. Although seemingly similar in form, the time-modulated Hamiltonian $H_{mod}(t) = a(t)H_{in} + b(t)H_{out}$ is completely different in essence from the stationary superposition Hamiltonian $H = a_w H_{in} + a_p H_{out}$. Here we leave it an open question for the readers to judge which way is closer to the true physics of these historical and modern MZI problems.

## V. Conclusion

In summary, we have used the standard quantum mechanical formalism to revisit and evaluate several classical MZI experiments used to test wave-particle duality of microscopic particles. We find that they are standard quantum mechanical scattering problem and the outcome wave function depends on the incident wave function and the Hamiltonian of each specific MZI. In particular, we have found that the classical delayed-choice experiments can be described by a time-modulated Hamiltonian $H_{mod}(t) = a(t)H_{in} + b(t)H_{out}$ describing the "in" and "out" status of BS2. As a result of this modulation, the final output wave function is also synchronically time-modulated, so that the outcome of measurement behaves as if they are in the superposition of the wave and particle quantum state when many events over time accumulation are counted and averaged. In comparison, the conventional wisdom holds that this is a quantum eigen-problem described by the stationary Hamiltonian $H = a_w H_{in} + a_p H_{out}$, and the solution yields a quantum superposition state as

$\psi = a_w \psi_{wave} + a_p \psi_{particle}$. Although these two interpretations yield the same averaged outcome of measurement in terms of wave (interference fringe pattern) and particle (which-way information distinguishability) behavior of quantum particles, the underlying physics is very much different. We expect that our new analysis over the delayed-choice MZI schemes may stimulate more studies over the conceptual basis of quantum physics, in particular, the wave-particle duality and orthodox Copenhagen principle of complementarity.

**Acknowledgement**

This work is supported by the 973 Program of China at No. 2013CB632704, the National Natural Science Foundation of China at Nos. 11434017, and financial support from the University of Chinese Academy of Sciences.

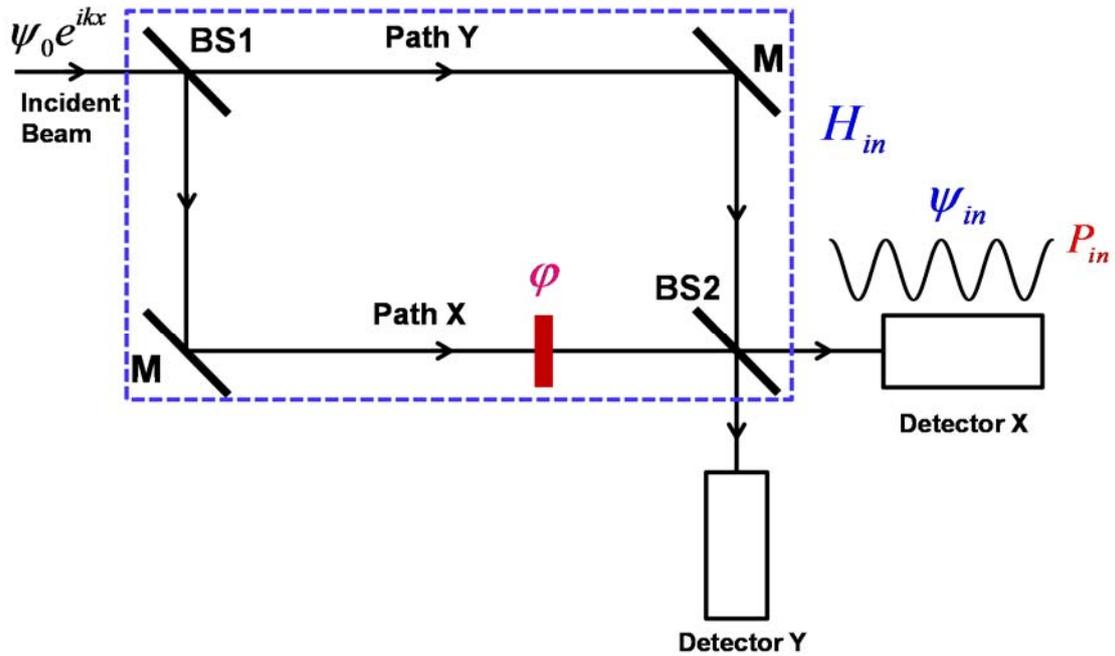

**Fig. 1.** Schematic setup based on the Mach-Zehnder two-arm interferometer used to probe the wave behavior of quantum particles. The blue dotted rectangular box is consisting of two beam slitters (BS1 and BS2), two mirrors (M), and a phase delay ($\varphi$). The quantum scattering system is described by the Hamiltonian $H_{in}$, which works upon the incident wave function $\psi_0 e^{ikx}$ and gives rise to the output wave function $\psi_{in}$. The signal recorded by the detector $x$ varies periodically with respect to the phase delay, remarking the wave nature of the quantum particle.

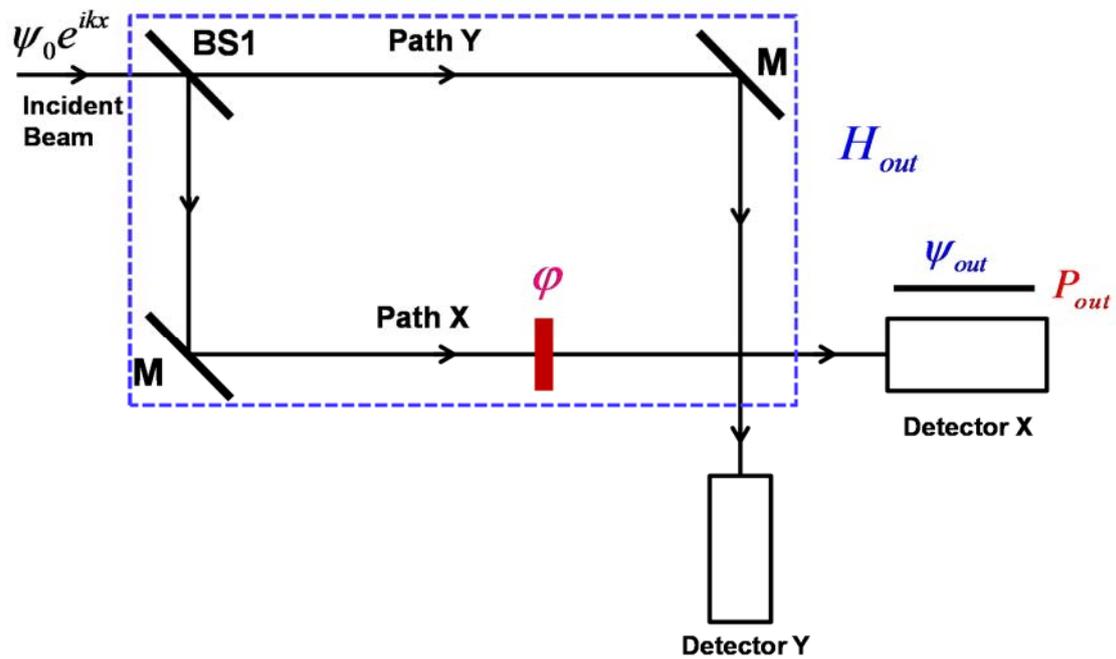

**Fig. 2.** Schematic setup based on the Mach-Zehnder two-arm interferometer used to probe the particle behavior of quantum particles. The blue dotted rectangular box is consisting of one beam slitter (BS1) but without the second splitter BS2, two mirrors (M), and a phase delay ($\varphi$). The quantum scattering system is described by the Hamiltonian $H_{out}$, which works upon the incident wave function $\psi_0 e^{ikx}$ and gives rise to the output wave function $\psi_{out}$. The signal recorded by the detector $x$ determines which-path information of the quantum particle, and is a constant with respect to the phase delay.

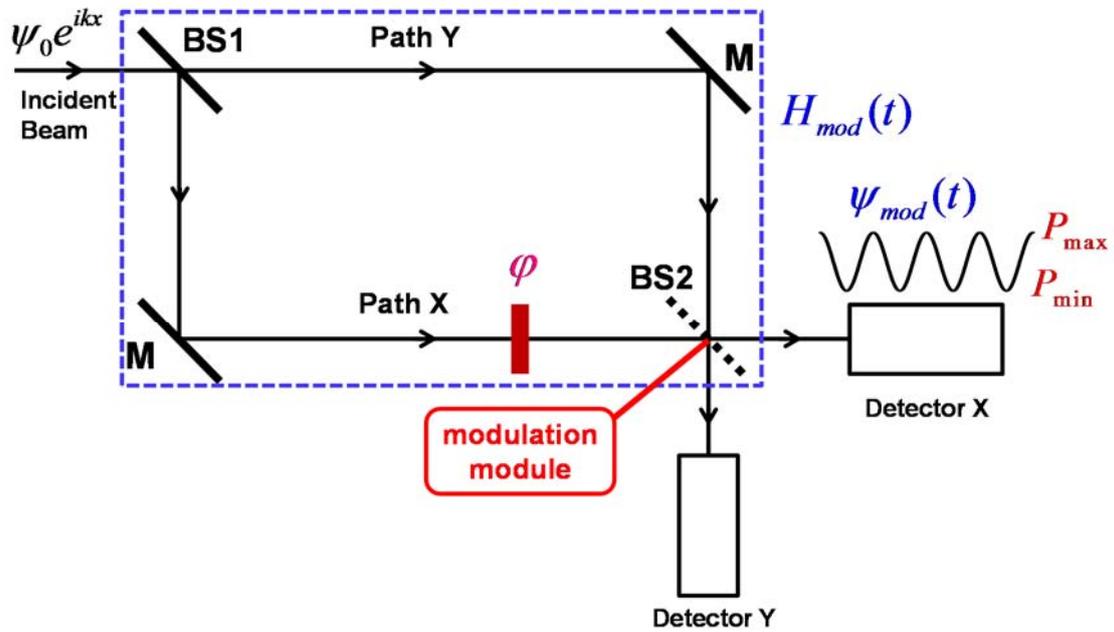

**Fig. 3.** Schematic setup of delayed-choice experiment based on the Mach-Zehnder two-arm interferometer. The blue dotted rectangular box is consisting of the first beam slitter BS1, two mirrors (M), a phase delay ($\varphi$), and the second beam splitter BS2 connected with modulation module to control its insertion into and removal from the path. The quantum scattering system is described by the time dependent Hamiltonian $H_{mod}(t)$, which works upon the incident wave function $\psi_0 e^{ikx}$ and gives rise to the output time-dependent wave function $\psi_{mod}(t)$. The signal recorded by the detector $x$ varies in time with a complicated behavior that can be described by the superposition of the wave and particle behavior of the quantum particle examined in Fig. 1 and Fig. 2, respectively.